\documentclass[12pt]{article}

\newcommand{\blind}{1}

\usepackage{authblk}
\usepackage{amsmath, amsthm, amssymb, bm}
\usepackage{graphicx, psfrag, epsf, subcaption}
\usepackage{enumerate}
\usepackage{natbib}
\usepackage{textcomp}
\usepackage[hyphens]{url} 
\usepackage{verbatim}

\usepackage{subfiles} 
\usepackage{dsfont}
\usepackage{xcolor} 
\usepackage{hyperref}
\hypersetup{%
	colorlinks = false
}

\usepackage{verbatim}

\usepackage[utf8]{inputenc}

\usepackage{float} 

\usepackage{listings}
\usepackage{adjustbox}
\usepackage{multirow} 

\usepackage{lineno} 

\usepackage[ruled,vlined]{algorithm2e}

\newcommand{\F}{{\mathcal F}}

\newtheorem{thm}{Theorem}

\newtheorem{assumption}{Assumption}[section]

\begin{document}
	
	\def\spacingset#1{\renewcommand{\baselinestretch}%
		{#1}\small\normalsize} \spacingset{1}

	\if1\blind
	{
		\title{\bf Evaluating real-time probabilistic forecasts with application to National Basketball Association outcome prediction}
		
		\author[1]{Chi-Kuang Yeh\thanks{chi-kuang.yeh@uwaterloo.ca}}
		\author[1]{Gregory Rice}
		\author[12]{Joel A. Dubin}
		\affil[1]{Department of Statistics and Actuarial Science, University of Waterloo, Waterloo, ON, Canada}
		\affil[2]{School of Public Health and Health Systems, University of Waterloo, Waterloo, ON, Canada}
		\setcounter{Maxaffil}{0}
		\renewcommand\Affilfont{\itshape\small}
		\maketitle
	} \fi
	
	\if0\blind
	{
		\title{ \bf Evaluating real-time probabilistic forecasts with application to National Basketball Association outcome prediction}
		
		\author[1]{  }
		\setcounter{Maxaffil}{0}
		\renewcommand\Affilfont{\itshape\small}
		\maketitle
	} \fi
	
	\begin{abstract}
		Motivated by the goal of evaluating real-time forecasts of home team win probabilities in the National Basketball Association, we develop new tools for measuring the quality of continuously updated probabilistic forecasts. This includes introducing calibration surface plots, and simple graphical summaries of them, to evaluate at a glance whether a given continuously updated probability forecasting method is well-calibrated, as well as developing statistical tests and graphical tools to evaluate the skill, or relative performance, of two competing continuously updated forecasting methods. These tools are studied by means of a Monte Carlo simulation study of simulated basketball games, and demonstrated in an application to evaluate the continuously updated forecasts published by the United States-based multinational sports network ESPN on its principle webpage {\tt espn.com}. This application lends statistical evidence that the forecasts published there are well-calibrated, and exhibit improved skill over several na\"ive models, but do not demonstrate significantly improved skill over simple logistic regression models based solely on a measurement of each teams' relative strength, and the evolving score difference throughout the game.
	\end{abstract}
	
	\noindent%
	
	\spacingset{1.45} 
	
	
	\section{Introduction}\label{sec-1}

	Probabilistic predictions and forecasts are ubiquitous in modern society, and many individuals consider, and make decisions based on, such forecasts on a routine basis. For example, in the United States probability of precipitation forecasts became widely publicly available starting in the late 1960's, and are now a critical factor in countless people's daily decisions \citep{nap-2006,murphy-1998}. Over time the number and scope of probabilistic forecasts readily accessible to the public has increased at a steady pace, and now covers prediction of phenomena ranging from sports \citep{silver-2019}, to politics \citep{erickson-2012}, to medicine \citep{spiegalhalter-1986}, to geology \citep{usga-2015}, among many other, some more exotic \citep{rowe-2018}, areas.
	
\begin{figure} [ht!]
		\centering
		\includegraphics[width=\textwidth]{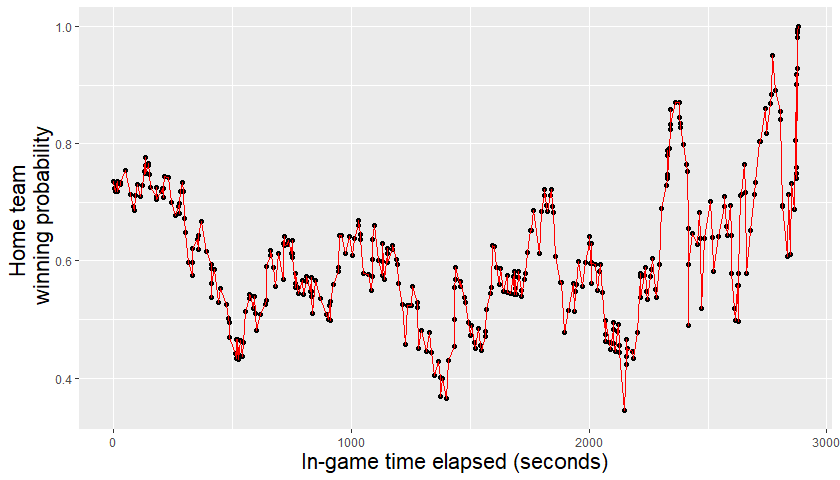}
		\caption{Real-time probabilistic forecasts as a function of the in-game time published on {\tt{espn.com/nba}} \citep{nba-espn} from a November 8, 2018 game in which the Los Angeles Lakers hosted the Minnesota Timberwolves. The black points are the points at which ESPN updated the home team winning probability forecasts due to events occurring in the game, and the red line is the linear interpolation between these forecasts to produce a real-time probabilistic forecast curve.} \label{fig-1game}
	\end{figure}

	Many such forecasts are made initially well before the event in question occurs, and are then continuously updated as new information becomes available. The example that we focus on throughout this paper is basketball game outcome prediction in the National Basketball Association (NBA). Websites like {\tt espn.com}, the main web page of the United States-based multinational sports network, ESPN, publish and update in real-time probabilistic forecasts of the home team winning for each NBA game played. Although the method by which ESPN produces these forecasts is largely proprietary, ostensibly initial probability forecasts of the home team winning are constructed based on information that is available before the game starts, e.g. the usual home court advantage in the NBA, relative team strength, player injuries, etc., and then after the game commences and progresses these forecasts are updated with new information such as the score, game time remaining, ball possession, fouls, in game player injuries, etc. The resulting probabilistic forecasts and their fluctuations may be viewed as a curve that is a function of the in-game time; see Figure \ref{fig-1game} for an example. Such curves arise in any similar continuously updated probabilistic forecasting task, and are evidently not unique to basketball game outcome prediction; see e.g. \cite{silver-politics}.

	Natural questions to ask when faced with any probabilistic forecast, including those that are continuously updated, are ``are these forecasts accurate?" and ``could these forecasts be improved upon?". Following the seminal work of \cite{murphy-1987} and \cite{Murphy-1992} on evaluating the quality of probabilistic forecasts in meteorology, evaluating a method for producing probabilistic forecasts is often broken into the tasks of measuring its {\it calibration} and {\it skill}. A model is deemed well-calibrated if its forecasts are compatible with the observed outcomes. In other words, a model that predicts an outcome with a given probability is well-calibrated if the relative frequency that the outcome occurs matches the probability forecast in the long run. A model is deemed to have higher skill than a competing model if its predictions are ``sharper" or ``more concentrated" than its competitor. For example, a model that forecasts the probability of a rainy day in New York City on a given day with the long run background rate of rainy days (which happens to be about 33.1\% for New York City), will in the long run be well-calibrated, but has less skill than a model, perhaps based on more complete weather data, that correctly predicts rainy days with probabilistic forecasts of zero and one. Excellent reviews and more in-depth discussions of these concepts can be found in \cite{gneiting-2007} and  \cite{gneiting-2014}.

	The goal of this paper is to develop simple, easily interpretable, tools for evaluating the calibration and skill of methods to produce continuously updated probabilistic forecasts, and to apply these methods to evaluate the probabilistic forecasts pertaining to NBA basketball game outcome prediction published on ESPN. In terms of evaluating model calibration of probabilistic forecasts, standard tools are reliability diagrams and calibration plots, in which outcome frequencies are plotted against binned forecast probabilities; see \cite{gneiting-2007}. Below we show how such curves can be extended in the continuously updated case to calibration surfaces, and how such surfaces can be summarized to show at a glance whether a given method is well-calibrated. In order to evaluate the relative skill of one continuously updated forecasting model against another, we employ the method of \cite{lai-2011} to construct confidence intervals for the average loss difference measured by the Brier score \citep{brier-1950} between the two models at given time points throughout the updating process. Estimating these intervals pointwise can be used to construct a simple graphical summary of the relative skill of one model versus another. In order to measure the cumulative statistical significance of differences observed in such a plot, we develop a new significance test for the skill-differences aggregated across time based on a novel large sample result for estimating continuous loss difference curves.
	
	For the purpose of demonstrating these methods and evaluating ESPN's forecasts, we introduce a number of ``competing" continuously updated forecasting methods for basketball outcome prediction. Some are designed to be ``straw men" for the purpose of demonstration, whereas others are based on logistic or probit generalized linear models making use of in-game information such as the score difference. We show using our methods that ESPN's model is generally well-calibrated, and exhibits significantly better skill than some na\"ive models, although it does not demonstrate superiority over relatively simple logistic regression models based on the score difference and relative team strength alone.

	The rest of the paper is organized as follows. Section \ref{sec-2} introduces the details of the ESPN forecasting data that we consider, as well as some competing forecasting methods that we develop and use for the purpose of comparison. Section \ref{sec-calib} discusses the construction of calibration surfaces for such forecasts, as well as simple graphical summaries of these surfaces. Section \ref{sec-method} explains the proposed methods to evaluate the relative skill of two sets of real-time probabilistic forecasts. A Monte Carlo simulation study of these methods is given in Section \ref{sec-sim}. A detailed comparison of the ESPN forecasts as well as those of the proposed models is given in Section \ref{sec-app}. Technical details are provided in Appendix \ref{sec-appendix} following these sections.
	
	\section{Motivating Data and Competing Models } \label{sec-2}
	

	The specific data that we consider and that motivates this work are play-by-play records and real-time probabilistic forecasts of NBA regular season games downloaded from {\tt{espn.com/nba}} (\cite{nba-espn}). The NBA is a major professional basketball league, which is often referred to as one of the ``Big Four" professional sports leagues in North America. Since 2004, except for the lockout season in 2011 and the COVID-19-influenced season of 2020, the NBA is comprised of $30$ teams, with each team playing a schedule of $82$ games in the regular season.
	
	Starting in the 2017 - 2018 NBA season, ESPN Analytics began providing real-time in-game probabilistic forecasts of the home team winning for each NBA game played; an example of the forecasts from one game is shown in Figure \ref{fig-1game}. The data available from ESPN are quite rich, including real-time information about details such as substitutions, fouls, and ball possession. We consider here only a subset of these data that includes the real-time probabilistic forecasts provided by ESPN, as well as the evolution of the score throughout the game, for the 2017-2018 and 2018-2019 seasons. These data are updated each time there is an ``event" in the game, which includes primarily score changes, fouls, and changes of possession. A typical game features between 460-480 events.
	
	We excluded a small portion of these data from our analysis due to two issues. Quite often, multiple events will occur at the same instant in a game. One of the main examples that contributes to this is multiple players substituting at the same time. Although these events are all logged at the same time point, they occur in the dataset in an ordered sequence. The forecasted probabilities published by ESPN during such an event are typically contingent on this order. Therefore, we simply average the forecasts together in such a scenario to produce a probabilistic forecast at that instant.  We also tried a number of other ways to handle this situation, such as using the first or last probabilistic forecast among the events recorded, and the difference in the results was negligible.
	
	The second issue is due to games that go to overtime. If two teams' scores are tied at the end of the 48-minute regulation game time, the teams will play an extra five-minute overtime period. For such games we remove the overtime period from the analysis, and only consider probabilistic forecasts up to the end of the game so that they are comparable to those that arise from games that did not go to overtime. Overtime games represent slightly less than 10\% of the total games. Additionally, a small number of data points were discarded due to evident defects or excessive missing values.
	
	The remaining data that we analyze are summarized in Table \ref{tb:data-quality}, and in each season there are more than $1100$ games with a total of over 350,000 play-by-play records available. Below we use the data from the 2017-2018 season as training data for our own models, and then we produce and evaluate forecasts for the 2018-2019 season.

	\begin{table}
		\centering
		\begin{tabular}{lllllll}
			\hline
			Season                   & Mode      & Games & Events & Max. events & Min. events & Avg number of events  \\
			\hline
			\multirow{3}{*}{17 - 18} & Raw       & 1158  & 530032 & 606       & 234       & 457.7133   \\
			& Selected    & 1137  & 517983 & 572       & 240       & 455.5699   \\
			& Processed   & 1137  & 354749 & 375       & 173       & 312.0343   \\
			\hline
			\multirow{3}{*}{18 - 19} & Raw       & 1229  & 583443 & 700       & 124       & 474.7299   \\
			& Selected    & 1213  & 572546 & 598       & 366       & 472.0082   \\
			& Processed & 1213  & 396991 & 385       & 241       & 327.2803  \\
			\hline
		\end{tabular}
		\caption{Summary of the data obtained from ESPN \citep{nba-espn} from the 2017-2018 and 2018 - 2019 NBA regular seasons. Raw counts represent the total number of games that ESPN provides probability forecasts for. Selected refers to those games that do not contain errors or missing values. Processed represents the data after averaging out multiple events recorded at the same game time.}
		\label{tb:data-quality}
	\end{table}
	
	Letting $N$ denote the total number of games with forecasts that we wish to evaluate, so $N=1213$ when we consider the 2018-2019 forecasts, the data may then be denoted as $\hat{p}_i^{ESPN}(t)$, $1\le i \le N$, $t \in [0,1]$ representing the probabilistic forecasts of the home team winning in the $i'th$ game at intra-game time $t$. We assume that the game time parameter $t$ is normalized to be between zero and one so that it represents the proportion of the game complete. Although these forecasts are only available when events occur, due to the fact that events are very dense throughout the game, we simply interpolate these forecasts linearly to produce full probability forecast curves over $[0,1]$, which also makes them more comparable from one game to the next. This is illustrated in Figure \ref{fig-1game}. We also consider the data $H_i(t)$ and $A_i(t)$, $1\le i \le N$, $t \in [0,1]$, denoting the home team score and away team score in the $i'th$ game, respectively, at proportion $t$ of the game. In our analysis below we frequently make use of the score difference $ScD_i(t) = H_i(t) - A_i(t)$, $1\le i \le N$, $t \in [0,1]$.
	
	The goal of the methods that we develop below is to evaluate the quality of the forecasts $\hat{p}_i^{ESPN}(t)$, $1\le i \le N$, $t \in [0,1]$. To do this we also develop a number of benchmark models that are used for the purpose of comparison.

	
	\subsection{Benchmark models for predicting NBA game outcomes} \label{sec2-2}
	
	We use the following notation below. We let $Y_i$ denote the indicator random variable that the home team wins in the $i'th$ game, so that $Y_i=1$ if the home team wins the $i'th$ game, and $Y_i=0$ if the home team loses the $i'th$ game. We are interested in forecasting or estimating the probability $p_i(t)$ that the home team wins, given the information up to time $t$ in the game, so that
	
	$$
	p_i(t) = P(Y_i = 1 | \mbox{ all information up to time $t$ in game $i$}).
	$$

	A more formal definition of $p_i(t)$ is given in the Appendix \ref{sec-appendix}, but is omitted here to lighten the technical detail in the text.
	
	$\hat{p}_i^{ESPN}(t)$ is in principle an estimate (forecast) of $p_i(t)$. In order to evaluate the quality of these forecasts, we consider a number of competing benchmark models, progressing from na\"ive to more realistic. The most complicated covariates that we consider to build these models are the score difference $ScD_i(t)$, and some measure of the relative strength of the teams, which we term $RS_i$. There are a number of ways of evaluating the relative strength of teams, including using the Elo rating system \citep{elo-1978}, which has been extensively used to rate the strength of basketball teams(see \cite{silver-2014}, \cite{silver-2015}, and \cite{silver-2019}) and odds in betting markets. We use as a proxy of the relative team $RS_i = \hat{p}_i^{ESPN}(0)$, the pre-game probability of the home team winning as forecast by ESPN. We considered a number of alternate metrics to define $RS_i$, and found that generally the results and conclusions of the below analyses did not change significantly, and so we use this quantity as to avoid the development of other proxies of relative team strength.

	The benchmark models that we consider are listed below in order of most to least na\"ive. Some of these are based on generalized linear models (GLMs) for binary response data, such as logistic regression, and we use $g(\cdot)$ to denote the GLM link function, which in our case is either the logit or probit link, see e.g. Chapter 4 of \cite{mccullagh1989generalized}. All GLMs were fit using the {\tt R} programming language, specifically the {\tt glm} function in  the stats package, version 4.0.2, which uses iteratively reweighted least squares. For each model we used the 2017-2018 season data as training data, and then produced rolling forecasts on the 2018-2019 season data to compare to the ESPN forecasts.
	
	\par \textbf{C}oin-\textbf{F}lip (\textbf{CF}): $\hat{p}_i(t)=0.5$ for all $t \in [0,1]$. \\

	\par Historic \textbf{Home} team \textbf{W}in \textbf{P}robability (\textbf{HomeWP}): $\hat{p}_i(t)=0.593$, which represents the frequency at which the home team won over the course of the NBA regular season games from 2008-2017.

	\par \textbf{P}re-\textbf{g}ame \textbf{R}elative \textbf{S}trength (\textbf{PgRS}) model: $\hat{p}_i(t)$ is forecast from the GLM
	$$
	g(p_i(t)) = \beta_{0}(t) +\beta_{1}(t) RS_i,~
	$$
	estimated pointwise at every game time $t$. \\

	\par \textbf{L}eading \textbf{S}tatus (\textbf{LS}) model: $\hat{p}_i(t)$ is forecast from the GLM
	$$
	g(p_i(t)) = \beta_{0}(t) +\beta_{1}(t) LS_i(t),
	$$
	where $LS_i(t)=1$ if $ScD_i(t)>0$, $LS_i(t)=0$ if $ScD_i(t)=0$, and $LS_i(t)=-1$ if $ScD_i(t)<0$.\\

	\par \textbf{Sc}ore \textbf{D}ifference without \textbf{Int}ercept (\textbf{ScDnoInt}) model: $\hat{p}_i(t)$ is forecast from the GLM
	
	$$
	g(p_i(t)) = \beta_{1}(t) ScD_i(t),
	$$
	estimated pointwise at every game time $t$.\\

	\par \textbf{Sc}ore \textbf{D}ifference (\textbf{ScD}) model: $\hat{p}_i(t)$ is forecast from the GLM
	$$
	g(p_i(t)) = \beta_{0}(t) +\beta_{1}(t) ScD_i(t),
	$$
	estimated pointwise at every game time $t$.\\
	
	\par {\textbf{P}re\textbf{g}ame \textbf{R}elative \textbf{S}trength and \textbf{L}eading \textbf{S}tatus (\textbf{PgRSLS}) model: $\hat{p}_i(t)$ is forecast from the GLM
		$$
		g(p_i(t)) = \beta_{0}(t) + \beta_{1}(t) RS_i + \beta_{2}(t) LS_i(t),
		$$
		estimated pointwise at every game time $t$. }\\
	
	\par \textbf{P}re\textbf{g}ame \textbf{R}elative \textbf{S}trength and \textbf{Sc}ore \textbf{D}ifference (\textbf{PgRSScD}) model: $\hat{p}_i(t)$ is forecast from the GLM
	$$
	g(p_i(t)) = \beta_{0}(t) + \beta_{1}(t) RS_i + \beta_{2}(t) {ScD_i(t)},
	$$
	estimated pointwise at every game time $t$. \\
	
\begin{figure}
		\centering
		
		\begin{minipage}{.5\textwidth}
			\centering
			\includegraphics[width=7cm]{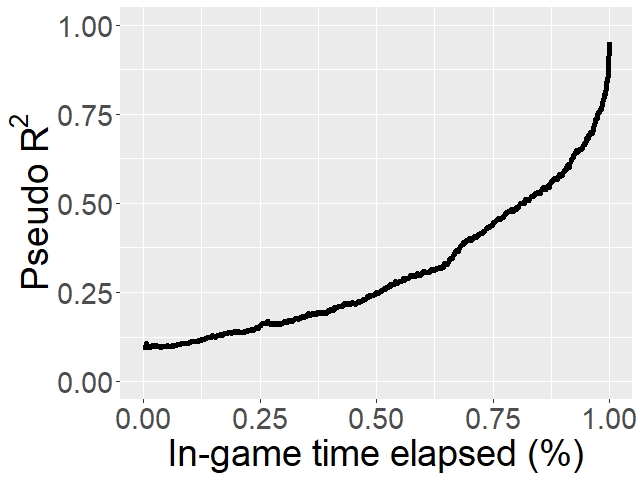}
		\end{minipage}%
		\begin{minipage}{.5\textwidth}
			\centering
			\includegraphics[width=7cm]{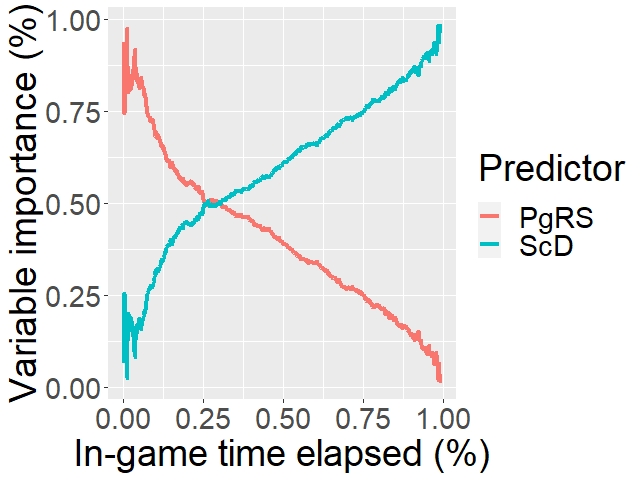}
		\end{minipage}
		\caption{The left hand panel shows the pseudo $R^2$ of the logistic regression model {\bf PgRSScD}, which uses the covariates pre-game relative strength and score difference, as a function of the game time. The right hand panel shows the variable importance of each covariate as it contributes to the pseudo $R^2$.  }\label{fig-varimp}
	\end{figure}

	We note that the GLMs with intercept terms are able to implicitly model the home team advantage, which refers to the phenomenon that in the NBA the home team tends to win a higher percentage of games than the away team, and so a model like {\bf ScDnoInt} could be expected to be poorly calibrated, at least at the beginning of the game. As mentioned, each GLM model is fit pointwise over the game-time parameter $t$. This allows that, for example, the effect, as determined by the models, of relative team strength, home team advantage, and score difference can evolve throughout the game. Diagnostic plots of the pseudo $R^2$ \citep{mcFadden-1973} and relative variable importance over the course of the game of the variables in our ``least na\"ive" model, {\bf PgRSScD}, are displayed in Figure \ref{fig-varimp}. It is clear from this figure that both models' predictions improve as the game progresses, evidently since ultimately the score difference covariate determines the winner, and also that the relative importance of the score difference versus team strength change inversely as the game progresses; relative team strength is the most important predictor early in the game, but becomes less important later in the game as the score difference becomes more informative.

	\section{Evaluating Calibration of Continuously Updated Forecasts}\label{sec-calib}
	
	We now turn to the task of evaluating the calibration for a given set of continuously updated forecasts $\hat{p}_i(t)$ with realized outcomes $Y_i$, $i=1,...,N$. As mentioned in the introduction, traditionally when evaluating such forecasts one often considers what are called {\it calibration plots} or {\it reliability diagrams}. A calibration plot is a plot of binned probabilistic forecasts against the conditional event frequency associated to forecasts in a given bin. Since well-calibrated forecasts should have that the event frequency match the forecast probability, calibration may then be measured by comparing these points against a 45-degree diagonal reference line. Large departures from this line thus indicate poor calibration (cf. \citep{Dawid-1986}, \citep{Murphy-1992} and \citep{Ranjan-2010}). We refer the reader to \citet{gneiting-2014} and the references therein for a more comprehensive discussion.
	
	One clear method then to check for calibration of continuously updated forecasts is to produce a calibration plot for each $t \in [0,1]$ based on the pairs $(\hat{p}_i(t),Y_i)$. While this is in essence what we propose, there are two main challenges in doing so. (1) Traditionally when producing calibration plots, the forecasted probabilities are binned into fixed bins, commonly by deciles. For example, often the event frequency corresponding to all forecasted probabilities between $[0,0.1)$ are compared to 0.05, similarly for [0.1,0.2) to $0.15$, and so on. With continuously updated forecasts, and as with the ESPN forecasts, it is typical that the forecasts fluctuate so that for certain time points $t$ the forecasts cluster around some fixed values, and are hence far from being uniformly distributed in such fixed bins. In the case of the ESPN forecasts, near the end of game the majority of forecasts are clustered around 0 and 1. Fixed bins will often have the problem that the forecasts within them are not uniformly distributed within the bin. (2) Having constructed calibration plots for each $t$, one must examine a large number of such plots to pinpoint if a method appears to be well-calibrated, or to diagnose if there are some subset of times $t$ at which the method is more or less calibrated than others. A simple summary of the many calibration plots produced would be useful.
	
	In order to address (1), we propose to use adaptive bins  in constructing the calibration plots. Specifically, for each $t$, suppose we wish to construct $M$ bins for the forecasts $\hat{p}_i(t)$. By calculating the ranked forecasts $\hat{p}_{(i)}(t)$, $i=1,...,N$, we may group them into $M$ bins so that $\hat{p}_{(i)}(t)$ is in bin $j$ if $\left(\lfloor N/M \rfloor(j-1) +1\right) \le i <  \lfloor N/M \rfloor j $. We denote the collection of $\hat{p}_i(t)'s$ in the $j$'th bin as $Bin_j$. In simpler terms, the forecasts at a given time $t$ are grouped into $M$ bins based on their rank. As a reference point or summary of the forecasts in the $j$'th bin, we use $\tilde{p}_j(t) = \mbox{Median}\left( \hat{p}_{(i)}(t), \;\; \left(\lfloor N/M \rfloor(j-1) +1 \right)\le i <  \lfloor N/M \rfloor j \right)$. A calibration plot at time $t$ is constructed then by comparing $\tilde{p}_j(t)$ to $\bar{Y}_j(t) = Average( Y_i , \mbox{ such that } \hat{p}_i(t) \in Bin_j)$. Letting $n_j$ denote the number of forecasts in $Bin_j$, a 95\% confidence interval for the mean of the events in $Bin_j$ could be constructed as
	
	$$
	\bar{Y}_j(t) \pm z_{1-{\alpha/(2M)}} \sqrt{\frac{\bar{Y}_j(t)(1-\bar{Y}_j(t))}{n_j}},
	$$
	where $z_\beta$ denotes the $\beta$ quantile of the standard normal distribution, and here we have applied a Bonferroni correction to the significance level according to the number of bins used $M$. $\alpha$ is typically taken to be 5\% so that 95\% confidence intervals are computed.
	
	The above interval is the standard confidence interval for the binomial proportion based on a normal approximation to the binomial random variable, as seen in most elementary statistics text books, and is often used to measure the uncertainty in calibration plots.  Often though in such a continuously-updated setting, as with the ESPN forecast data, the standard interval is poor since the event frequency may be very close to zero or one in some bins at a subset of time points $t$. As discussed at length in \cite{brown2001}, a confidence interval that is in general recommended as a replacement to the standard interval, and is more robust to event frequencies close to zero and one, is the Wilson interval \citep{wilson-1927}, which takes the form
	
	$$\frac{n_j \bar{Y}_j(t)+\kappa^{2} / 2}{n_j+\kappa^{2}} \pm \frac{\kappa n_j^{1 / 2}}{n_j+\kappa^{2}}\left(\bar{Y}_j(t)  (1-\bar{Y}_j(t))+\kappa^{2} /(4 n_j)\right)^{1 / 2},$$
	where $\kappa=z_{\alpha/(2M)}$. We have noticed significant improvements by using the Wilson interval in this setting due to event frequencies that approach zero and one near the end of the game, and so we use these intervals in calibration plot construction below. Additionally, so that no bins contain predominately zero or one probability forecasts, we discard all forecasts to produce a calibration plot at a given $t$, and corresponding events, if $\hat{p}_i(t)<0.005$ or $\hat{p}_i(t)>0.995$, and we analyze those forecasts for calibration separately; see Table \ref{extr-tab}.
	
\begin{figure} [ht!]
		\centering
		\begin{subfigure}[b]{0.455\textwidth}
			\centering
			\caption[]%
			{{\small}}
			\includegraphics[width=\textwidth]{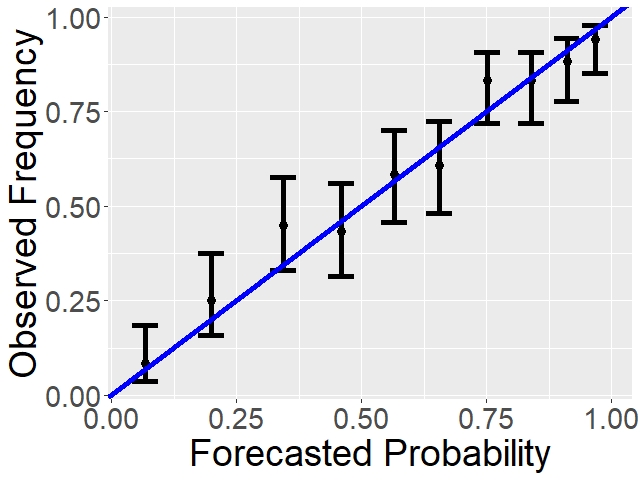}
			\label{fig-cali-espn-t05}
		\end{subfigure}
		\hfill
		\begin{subfigure}[b]{0.500\textwidth}
			\centering
			\caption[]%
			{{\small}}
			\includegraphics[width=\textwidth]{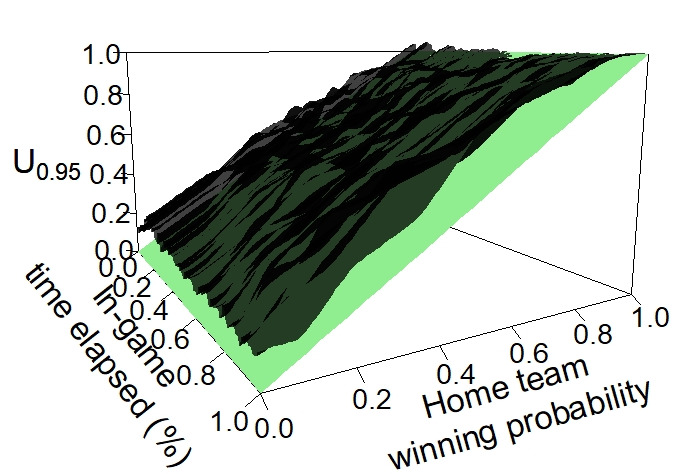}
			\label{fig-cali-surface}
		\end{subfigure}
		\caption{The left hand panel shows a calibration plot of the ESPN forecasts at time point $t=0.5$. The right hand panel shows an upper calibration surface along with reference plane $f(t,p)=p$ for the continuously updated ESPN forecasts obtained by interpolating the upper bounds of each confidence interval of the calibration plots across $t\in [0,1]$. } \label{fig-veri-plot}
	\end{figure}

	A calibration plot may then be made by plotting the bin references $\tilde{p}_j(t)$ against the above confidence intervals, and comparing these intervals to the reference line $y=x$. The method is deemed well-calibrated at the given significance level if the reference line generally goes through each interval. An example of this is shown based on the ESPN forecasts for $t=0.5$  using $M=10$ bins in the left hand panel of Figure \ref{fig-cali-surface} at the 95\% confidence level. By linearly interpolating the upper bounds of these intervals across both the reference proportions $\tilde{p}_j(t)$ as well as $t$, we may construct an ``upper calibration surface", $U_{1-\alpha}(t,p)$. The upper 95\% calibration surface $U_{0.95}(t,p)$ is displayed in the right hand panel of Figure \ref{fig-cali-surface} for the ESPN forecasts from the 2018-2019 season. A lower surface $L_{1-\alpha}(t,p)$ may be similarly constructed by linearly interpolating the lower bounds. A continuously updated forecasting method may then be deemed well-calibrated at a given significance level if the reference plane $f(t,p)=p$, for $t,p \in [0,1]$ is contained between both surfaces.

	Although such surface plots are informative, it can be challenging to infer quickly based on these plots whether a given method appears to be calibrated. In order to produce a more easily interpretable summary of such calibration surface plots, we instead consider a plot for each $t$ of the minimum distance over $p$ between the reference plane $f(t,p)=p$, and the upper and lower calibration surfaces. Specifically, we consider plots of the functions  $U^{min}_{1-\alpha}(t)= \min_{1 \le j \le M} U_{1-\alpha}(t,\tilde{p}_j(t)) - \tilde{p}_j(t)$ and  $L^{max}_{1-\alpha}(t)= \max_{1 \le j \le M} L_{1-\alpha}(t,\tilde{p}_j(t)) - \tilde{p}_j(t)$ against $t$. If the upper and lower confidence surfaces contain the reference plane $f(t,p)=p$, then $U^{min}_{1-\alpha}(t)$ should always lie above zero, and $L^{max}_{1-\alpha}(t)$ should always lie below zero. Points with respect to $t$ at which this does not hold can be used to identify times for which a given method is not well-calibrated. We note here that in this case, due to the high degree of fluctuations in continuously updated probabilistic forecasts for basketball prediction, we have found it also useful to aid in the interpretation of these plots to smooth them with respect to $t$ using a simple moving average smoother over 5\% of the game times. The conclusions drawn from these plots change little for different values of the window width.
	
\begin{figure} [ht!]
		\begin{subfigure}[b]{0.475\textwidth}
			\centering
			\caption[]%
			{{\small}}
			\includegraphics[width=\textwidth]{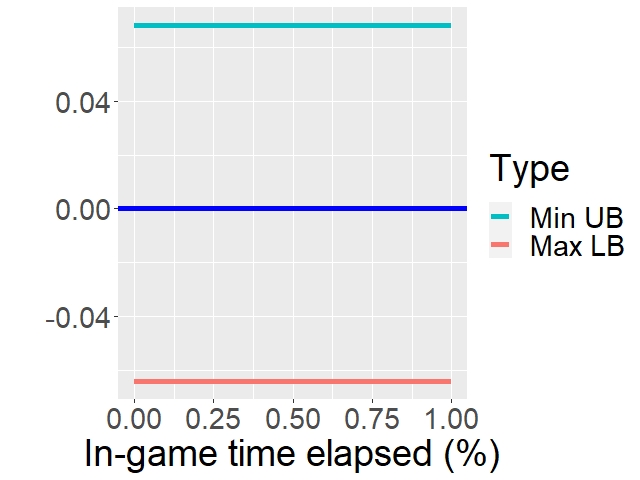}
			\label{fig-mm-homeWP}
		\end{subfigure}
		\hfill
		\begin{subfigure}[b]{0.475\textwidth}
			\centering
			\caption[]%
			{{\small}}
			\includegraphics[width=\textwidth]{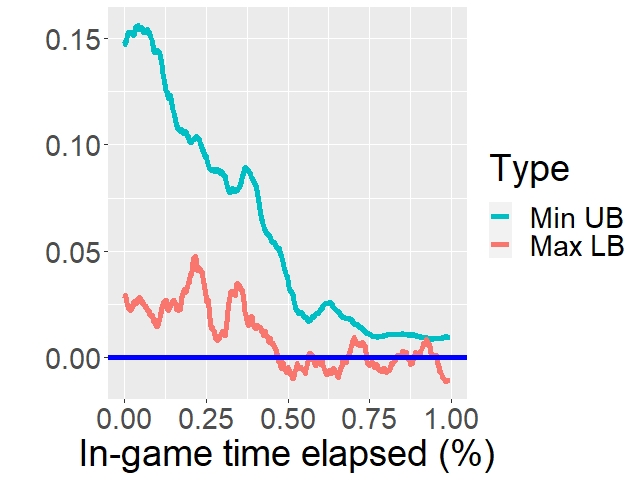}
			\label{fig-mm-SDnoINT}
		\end{subfigure}
		\hfill
		\begin{subfigure}[b]{0.475\textwidth}
			\centering
			\caption[]%
			{{\small}}
			\includegraphics[width=\textwidth]{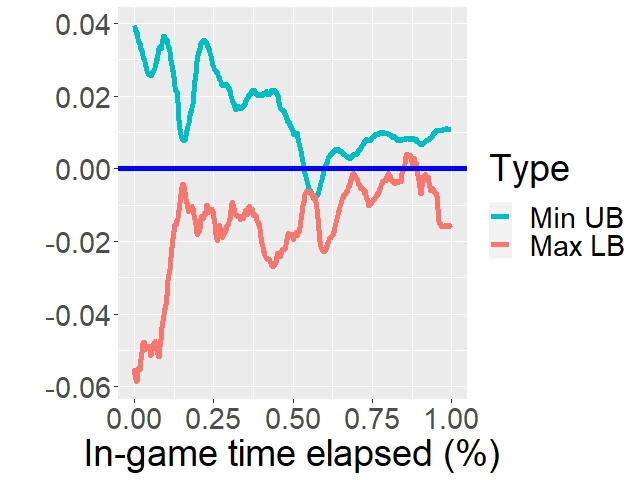}
			\label{fig-mm-espn}
		\end{subfigure}
		\hfill
		\begin{subfigure}[b]{0.475\textwidth}
			\centering
			\caption[]%
			{{\small}}
			\includegraphics[width=\textwidth]{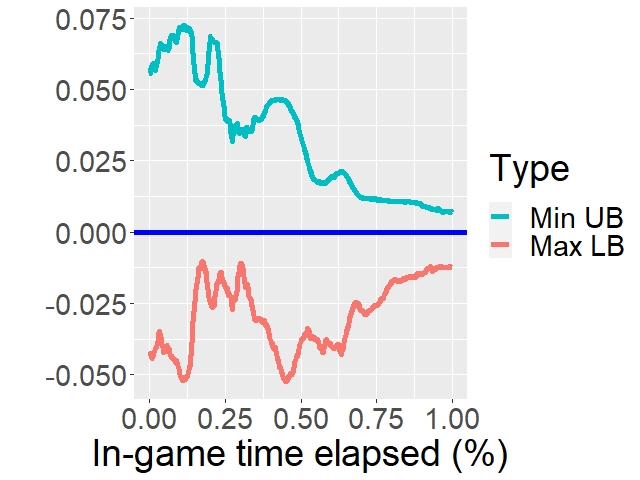}
			\label{fig-mm-logit}
		\end{subfigure}
		\caption{Plots of $U^{min}_{0.95}(t)$ and  $L^{max}_{0.95}(t)$ against $t$ based on $M=10$ bins: Methods (a) historic home team win probability ({\bf HomeWP}); (b) logit model using Score difference without intercept ({\bf ScDnoInt}); (c) ESPN forecasts; (d) logit model based on Pre-game relative strength and Score difference ({\bf PgRSScD}).} \label{fig-mmplots}
	\end{figure}

	These summary plots calculated based on the ESPN forecasts as well as from the benchmark models {\bf HomeWP}, {\bf ScDnoInt}, and {\bf PgRSScD} using the logit link function are shown in Figure \ref{fig-mmplots}. From these we see that the na\"ive model {\bf HomeWP}, which predicts that the home team will win each game at all times $t$ with the prior 10-year historic win rate of home teams in the NBA, is well-calibrated, as expected. Similar plots (not shown) for the method {\bf CF}, which simply predicts that the home team will win with probability 50\%, show that this method is not well-calibrated. Considering this plot for the method {\bf ScDnoInt} (Panel (b) in Figure 4), we see that the forecasts are poorly calibrated at the beginning of the game, but the calibration improves towards the end of the game. This is expected since the corresponding logit model is taken to be free of an intercept term, and so the model is unable to capture the home team advantage which should force the forecast probabilities to favour the home team at the beginning of the game. Both the ESPN forecasts and those from the model {\bf PgRSScD}, which incorporates team strength as well as the score difference, demonstrated generally good calibration for all game times.
	
	A summary of the outcomes corresponding to games in which a probability forecast at some intra-game time exceeded 0.995 or was less than 0.005 is given in Table \ref{extr-tab} for the ESPN forecasts as well as for the forecasts based on the models {\bf PgRSScD} and {\bf ScDnoInt}. The empirical home team win rate closely matched these thresholds in each case suggesting that the forecasts for each of these methods are reasonably well-calibrated at these extremal probability forecast levels.

	\begin{table} \label{tb-disregard-phats}
		\centering
		\begin{tabular}{lllllll}
			& \multicolumn{2}{c}{ESPN } & \multicolumn{2}{c}{PgRSScD} & \multicolumn{2}{c}{ScDnoInt}  \\
			\hline
			$\hat{p}_i(t)$ for some $t$          & $>0.995$     & $<0.005$            & $>0.995$     & $<0.005$               & $>0.995$     & $<0.005$                          \\
			\hline
			Total Games    & 555    & 343              & 591    & 358              & 589    & 374                          \\
			Home team wins   & 553      & 1                & 588      & 0                & 587      & 0                            \\
			Proportion     & 0.9964 & 0.0029           &0.9949 & 0.0000           & 0.9966 & 0.0000                       \\
			\hline
		\end{tabular} \caption{The number of games in which a probability forecast exceeded 0.995 or was less than 0.005  at some point for each of the ESPN, PgRSScD with logit link, and the ScDnoInt with logit link forecasts, as well as the number and proportion among these games in which the home team won.}\label{extr-tab}
	\end{table}

	\section{Evaluating Skill of Continuously Updated Forecasts} \label{sec-method}
	
	\subsection{Pointwise confidence intervals measuring the skill difference between competing methods} \label{sec3-1}
	
	As described in the introduction, the skill of a probabilistic forecasting method generally refers to its ``sharpness" or acuity relative to a competing or benchmark method. Formally this can be measured by defining a loss function, or scoring rule, used to measure the accuracy of a given probabilistic forecast based on the realized events. Perhaps the most frequently used loss function is $L(a,b)=(a-b)^2$ and defines the Brier score (\cite{brier-1950}). We use this loss function below, but the following results generalize to any loss function that has a linear equivalent, which means that (1) $L'(x,b)$ is linear in $x$, and (2) $L'(x, b) - L(x,a)$ does not depend on $x$. This includes Kullback Divergence \citep{kullback-1951}, and Good's log-score, among others; see \cite{lai-2011}, \cite{gneiting-2007} and \cite{bickel-2007}.
	
	Following the work of \cite{lai-2011}, ideally any method to forecast $p_i(t)$ would satisfy that $L(p_i(t),\hat{p}_i(t))$ is small, and further when averaged over all forecasts would minimize
	
	\begin{equation}\label{eq-loss}
		L_{N}(t)=\frac{1}{N} \sum_{i=1}^{N} L\left(p_{i}(t), \hat{p}_{i}(t)\right).
	\end{equation}
	
	Since the underlying true probabilities $p_i(t)$ are unobservable, a sensible estimate of $L_N(t)$ is obtained by replacing these probabilities with their point estimates based on the realizations $Y_i$ to produce
	\begin{equation}\label{eq-Lhat}
		\hat{L}_N(t)=\frac{1}{N}\sum_{i=1}^N L(Y_i,\hat{p}_i(t)).
	\end{equation}
	
	The quantity $\hat{L}_N(t)$ captures the skill or ``sharpness" of the forecasting method at a given time point $t$ as described above, since a given forecasting method is ascribed generally lower losses, or higher scores, if $\hat{p}_i(t)$ is closer $Y_i$, the latter of which takes the values $0$ and $1$.
	
	Suppose we wish to compare two methods, call them method $A$ and method $B$, for producing continuously updated probabilistic forecasts. We denote such forecasts by $\hat{p}_i^A(t)$ and $\hat{p}_i^B(t)$, and we compare them based on the corresponding realized events $Y_i$, $1 \le i \le N$. This can be done by comparing their average losses defined in \eqref{eq-Lhat}. Specifically, we consider the function of $t$
	
	\begin{equation} \label{eq-hat-delta}
		\hat{\Delta}_N(t)=\frac{1}{N}\sum_{i=1}^N\left[L(Y_i,\hat{p}_i^A(t))- L(Y_i,\hat{p}_i^B(t)) \right],
	\end{equation}
	which can be viewed as an estimate of the true loss difference
	\begin{equation} \label{eq-delta}
		\Delta_N(t)=\frac{1}{N}\sum_{i=1}^N\left[L(p_i(t),\hat{p}_i^A(t))- L(p_i(t),\hat{p}_i^B(t))\right].
	\end{equation}
	
	Larger than zero values of $\hat{\Delta}_N(t)$ favour method $A$ at the given $t$, whereas negative values show favour for method $B$. In order to measure the statistical significance of any deviations of $\hat{\Delta}_N(t)$ from zero, we may view $\hat{\Delta}_N(t)$ as an estimator for $\Delta_N(t)$. By constructing suitable confidence intervals for $\Delta_N(t)$ based on $\hat{\Delta}_N(t)$, we may evaluate whether observed deviations suggest the superiority of one model over another, and further construct simple graphical summaries that illustrate the skill of one model compared to another as a function of $t$.
	
	To construct such confidence intervals, we first define the variance of $\hat{\Delta}_N(t)$ as
	
	\begin{align}\label{eq-sndef}
		s_N^2(t) = \frac{1}{N} \sum_{i=1}^N \delta_i^2(t) p_i(t)(1-p_{i}(t)),
	\end{align}
	where $\delta_{i}(t) = \left[L(1, \hat{p}_{i}^A(t))- L(0, \hat{p}_{i}^A(t))\right]- \left[L(1, \hat{p}_{i}^B(t))-L(0, \hat{p}_{i}^B(t))\right]$. The following result is proved in \cite{lai-2011} and stated for each $t\in [0,1]$. \\
	
	{\it Theorem 2, \cite{lai-2011}: } Suppose that for each $t\in [0,1]$ that $s_N^2(t)$ converges in probability to a positive constant as $N \to \infty$, and that the variables $A_i(t) = L(Y_i,\hat{p}_i^A(t))- L(p_i(t),\hat{p}_i^A(t))$ and  $B_i(t) = L(Y_i,\hat{p}_i^B(t))- L(p_i(t),\hat{p}_i^B(t))$ each form martingale difference sequences. Then for each $t$,
	
	$$
	\frac{\hat{\Delta}_N(t) - {\Delta}_N(t)}{s_N(t)} \stackrel{D}{\to} \mathcal{N}(0,1),
	$$
	where $\stackrel{D}{\to}$ denotes convergence in distribution, and $\mathcal{N}(0,1)$ denotes a standard normal random variable.

	The two main conditions of the above theorem are that (1) $s_N^2(t)$, the variance of $\Delta_N(t)$, should for large $N$ behave like a positive constant, and (2) that the forecast loss differences should behave like martingale difference sequences. The first condition can be thought of as a non-degeneracy condition-- this result only holds if the forecasts of the two methods to be compared do not coincide entirely. It is not valid for two methods that produce equivalent, or almost equivalent, forecasts. This is, in general, a reasonable assumption if the forecasts being compared are from entirely different models, or if one or both sets of forecasts to be compared come from unknown models, as with the ESPN forecast data, since it is unlikely in this case that they will produce forecasts that coincide. This assumption is in question when comparing forecasts of nested models, e.g. comparing two GLM models whose only difference is that a covariate is included or excluded. A more in-depth discussion of comparing forecasts from nested models, and the problems that arise from it, may be found in \cite{clark-2015}. Regarding condition (2), this is almost always satisfied when using a loss function with a linear equivalent and constructing genuine forecasting methods that must be based on available (past) information, rather than information from the unknown future, as discussed on page 2361 of \cite{lai-2011}, and so this condition should be thought of as mild.
	
	The above results suggest constructing a $100(1-\alpha)\%$ confidence interval for $\Delta_N(t)$ as
	\begin{equation}\label{eq-conf-int-diff}
		\hat{\Delta}_N(t)\pm z_{1-\alpha/2}\frac{s_N(t)}{\sqrt{N}}.
	\end{equation}
	
	Note that since $p_i(t)(1-p_i(t))$ in the definition of $s_N^2(t)$ is unobserved, we may replace it with the upper bound of $1/4$ in both (\ref{eq-sndef}) and (\ref{eq-conf-int-diff}) to obtain a conservative confidence interval. Plots as a function of $t$ of $\hat{\Delta}_N(t)$ and the corresponding conservative confidence intervals $\hat{\Delta}_N(t)\pm z_{1-\alpha/2}\frac{s_N(t)}{\sqrt{N}}$ can be used to evaluate the relative skill of one model compared to another. Points $t$ at which the associated confidence interval $1-\alpha$ confidence interval for $\Delta_N(t)$ do not contain zero indicate a significant improvement at the level $\alpha$ of the average loss of one method over another. Examples of plots of this form may be found in Figure \ref{fig-loss-models}, which we discuss in more detail below. We note again that due to the high degree of fluctuations in continuously updated probabilistic forecasts for basketball prediction, we have found it also useful to aid in interpretation of these plots to smooth them with respect to $t$ using a simple moving average smoother over 5\% of the game times.
	

	\subsection{ Functional tests to measure skill aggregated across $t$}\label{sec33}
	
	Although the above confidence intervals can be used to evaluate whether two methods exhibit similar or significantly different skill at any given time point $t$, it is often also of interest to evaluate whether two continuously updated models have approximately equal predictive power when discrepancies between them are aggregated across $t \in [0,1]$. For example, it might be that one method exhibits similar but somewhat better skill at each game time that when viewed in aggregate suggest the superiority of one model over another. Conversely, it is also possible one method exhibits apparently improved performance at a single game time $t_0$, although when viewed in aggregate across $t \in [0,1]$ this improvement may appear rather insignificant.

	To make this more precise, we formulate the null hypothesis of equal predictive power aggregated accross $t \in [0,1]$ of two methods as

	$$
	H_0: \; \|\Delta_N\|^2 = 0,
	$$
	
	where $\|\cdot\|^2$ is the standard squared $L^2$ norm of a function, so that
	$$
	\|f\|^2 = \int_{0}^{1}f^2(t)dt.
	$$
	
	The hypothesis $H_0$ posits then that the two methods to be compared exhibit approximately on average (across $t$) equal skill. Let
	
	\begin{align*}
		Z_N(t) &= \sqrt{N}\hat{\Delta}_N(t).
	\end{align*}
	A measure of global discrepancy across time between the two forecasting methods may be obtained by considering $\|Z_N\|^2$. In order to determine the large-sample properties of $Z_N$ that would inform determining appropriate significance levels and estimating $p$-values for tests of $H_0$ based on $\|Z_N\|^2$, we make use of the following result, which we state  rigourously and prove in Appendix \ref{sec-appendix}.
	
	%
	
	\begin{thm}\label{thm-test}
		Under $H_0$ and conditions analogous to those of Theorem 2 of \cite{lai-2011}, see Appendix \ref{sec-appendix} for details, there exists an infinite sequence of constants $\{\lambda_i, \; i\ge 1\}$ that satisfy
		
		$$
		\lambda_1 \ge \lambda_2 \ge \cdots \ge 0, \mbox{ and } \sum_{i=1}^{\infty} \lambda_i <  \infty,
		$$
		so that
		$$
		\|Z_N\|^2 \stackrel{D}{\to} \sum_{i=1}^{\infty} \lambda_i \chi_i^2(1),
		$$
		where $\chi_i^2(1)$, $i=1,2,\ldots$ are independent and identically distributed $\chi^2$ random variables with one degree of freedom. Moreover, the constants $\{\lambda_i, \; i\ge 1\}$ can be conservatively estimated by the eigenvalues of the function
		$$
		\hat{C}_{cons}(t,s) = \frac{1}{N} \sum_{i=1}^{N} [\hat{p}^A_i(t)-\hat{p}^B_i(t)][\hat{p}^A_i(s)-\hat{p}^B_i(s)].
		$$
		Namely with $\{\hat{\lambda}_i, \; i =  1,...,N\}$ defined so that $\hat{\lambda}_1 \ge \hat{\lambda}_2 \ge \cdots \hat{\lambda}_N \ge 0$ and  satisfying that there exist functions $\hat{\phi}_i(t)$, $i=1,...,N$, $t\in[0,1]$ with $\|\hat{\phi}_i\|^2 =1$, such that
		
		\begin{align}\label{cons-eig}
			\hat{\lambda}_i \hat{\phi}_i(t) = \int_{0}^{1} \hat{C}_{cons}(t,s)\hat{\phi}_i(s)ds,
		\end{align}
		then for any fixed $j\ge 1$, $P( \hat{\lambda}_j > \lambda_j) \to 1$ as $N\to \infty$.
		
	\end{thm}
	
	This result suggests a simple way of conducting an approximate and conservative test of the hypothesis $H_0$. \\
	
	Step 1: Evaluate $\|Z_N\|^2$.   \\
	
	Step 2: Estimate $\hat{C}_{cons}$, and the eigenvalues satisfying \eqref{cons-eig}. \\
	
	Step 3: Estimate the distribution of the random variable $Q_D = \sum_{i=1}^{D} \hat{\lambda}_i \chi_i^2(1)$ where $D$ is a large number (below we take $D=10$ and have found this choice generally adequate. This can be done easily using Monte-Carlo simulation, or using the numerical method of \cite{imhof-1961}. \\
	
	Step 4: Calculate an approximate and conservative $p$-value of the test of $H_0$ as $p= P(Q_D \ge \|Z_N\|^2)$. \\
	
	This $p$-value combined with the confidence intervals in \eqref{eq-conf-int-diff} allow for a detailed evaluation, both at particular game times $t$ and across all $t\in[0,1]$, of the relative skill of competing continuously updated methods.
	
	\section{Simulation} \label{sec-sim}
	\par
	Since many of the above methods are new when applied to continuously updated probabilistic forecasts, in this section we present the results of a small simulation study to investigate their respective performances, and illustrate their application in some controlled examples.
	\subsection{Basketball game simulation}\label{sec-sim-dgp}
	
	
	\par
	We now turn our attention generating data that resembles the NBA game data that we described in Section \ref{sec-2}. This entails generating random quantities that serve the role of the score difference and the initial relative team strength of the teams to play. Let $\left\{W_{i}(t), t \in[0,1]\right\}, i=1, \ldots, N$ be independent standard Brownian motions, where again $N$ is the total number of games. To represent relative team strength, we consider $RS_i \sim a \times Unif(-1,1) + c$, where $Unif(-1,1)$ denotes a uniform random variable on $[-1,1]$, and $a,c\in \mathbb{R}$ are constants that we use to calibrate the simulated data. With this defined, we model the score difference as $ScD_{i}(t)=tRS_{i} +W_{i}(t)$, and we hence define the indicator variables (analogous to the home team winning) $Y_i = 1$ if $ScD_i(1)>0$, and $Y_i=0$ otherwise. Namely, the score difference is modelled as a Brownian motion with drift determined by $RS_{i}$; positive $RS_i$ makes it easier for the home team to win, while negative $RS_i$ has the opposite effect. The constants $a$ and $c$ defining $RS_i$ were selected so that the home team win probability is  approximately $59\%$ in order to match the past-10-year historical home team win rate in the NBA. Similar and simple Brownian motion models for NBA scores have been extensively studied; see \cite{stern1994brownian}, \cite{gabel2012random}, and \cite{chen2018functional}.
	
	Under these settings, it is relatively straightforward to show that
	\begin{align*}
		p_{i}(t) &=P\left(Y_{i}=1 | \mbox{ all information up to time $t$ in game $i$}\right) = \Phi\left(\frac{Sc D_{i}(t)+RS_{i}(1-t)}{\sqrt{1-t}}\right),
	\end{align*}
	where $\Phi$ denotes the standard normal distribution function. We call this true probability function the ``{\bf Oracle}" model or probability below.
	
	\subsection{Models}
	\label{sec-sim-model}
	
	\par
	
	We first consider the problem of constructing two ``models" for this simulation experiment that have equal forecasting power and satisfy the null hypothesis $H_0$ of equal forecasting accuracy. This is achieved in this setting by perturbing the Oracle probability by independent random noise that has the same distribution in each model to be compared. We considered two such ``models" in our simulation experiments:
	
	\par The \textbf{Ora}cle model perturbed by standard \textbf{B}rownian \textbf{m}otion (\textbf{OraBM}):
	$$\hat{p}_{i}(t) = \Phi\left(\frac{ScD_{i}(t)+RS_{i}(1-t)+BM_i(t)}{\sqrt{1-t}}\right),$$
	where $\left\{BM_{i}(t), t \in[0,1]\right\}, i=1, \cdots, N$ are independent standard Brownian motions.
	
	\par
	The \textbf{Ora}cle model perturbed by time homogeneous \textbf{O}rnstein–\textbf{U}hlenbeck process (\textbf{OraOU}):
	$$\hat{p}_{i}(t) = \Phi\left(\frac{ScD_{i}(t)+RS_{i}(1-t)+OU_i(t)}{\sqrt{1-t}}\right),$$
	where $\left\{OU_{i}(t), t \in[0,1]\right\}, i=1, \cdots, N$ are independent time-homogeneous standard Ornstein–Uhlenbeck processes. We may generate the $OU_i(t)$ from the standard Brownian motion as follows: $OU_i(t) = exp(-t/2) BM_i(e^{t})$.
	
	We use the notation {\bf OraBM1} and {\bf OraBM2} (similarly {\bf OraOU1} and {\bf OraOU2}) to denote two probabilistic forecasts generated according to the above models with independent noise terms. Since the noise term in both forecasts are equal in distribution, they intuitively have equal predictive power and satisfy $H_0$. We compare these ``models", as well as the GLM models introduced in Section \ref{sec2-2}. To fit these GLM models in a way that is similar to our analysis of the ESPN forecast data, we first generate two independent ``seasons" following the description above with $N$ games in each season, one of which we refer to as training data, and the other we call the testing data. We fit the competing models introduced in Section \ref{sec2-2} on the training with the link function $g(\cdot)$ taken to be the Probit link, and then compare the model forecasts on the test data. We note that with the Probit link the model {\bf PgRSScD} is correctly specified in the sense that $\Phi^{-1}(p_i(t))$ is for each fixed $t$ a linear function of $RS_i$ and $ScD_i(t)$.
	
	\subsection{Results}
	\label{sec-sim-result}
	
	For each setting of $N=100, 250$ and $500$ and model comparison considered, we simulated data independently $1000$ times and applied the test of $H_0$ described for comparing forecast skill in Section \ref{sec33}. The results in terms of empirical rejection rates of $H_0$ are summarized for each comparison made in Table \ref{tbl-sim-rej}.
	\par
\begin{figure} [ht!]
		\centering
		\begin{subfigure}[b]{0.475\textwidth}
			\centering
			\caption[]%
			{{\small}}
			\includegraphics[width=\textwidth]{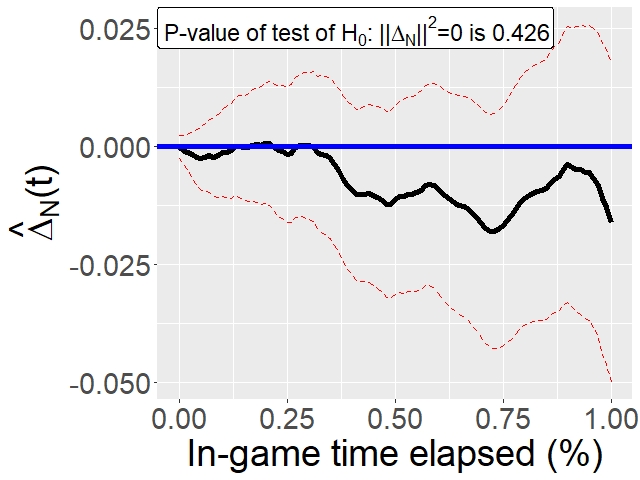}
		\end{subfigure}
		\hfill
		\begin{subfigure}[b]{0.475\textwidth}
			\centering
			\caption[]%
			{{\small}}
			\includegraphics[width=\textwidth]{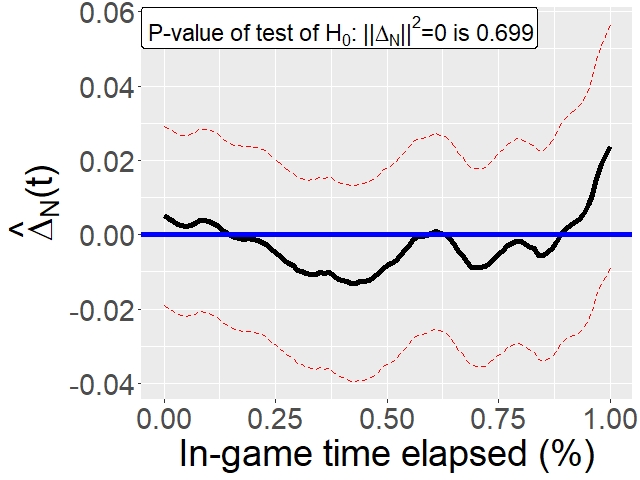}
		\end{subfigure}
		\caption{ Representative plots of $\hat{\Delta}_{1000}(t)$ with 95\% confidence intervals and approximate $p$-values for tests of $H_0$ for (a) {\bf OraBM1} versus {\bf OraBM2}; (b) {\bf OraOU1} and {\bf OraOU2}.}\label{fig-sim-noise}
	\end{figure}

	We first compare the oracle models perturbed by noise {\bf OraBM1} versus {\bf OraBM2}, and  {\bf OraOU1} versus {\bf OraOU2}. Figure \ref{fig-sim-noise}  gives a representative example of plots  as a function of $t$ of  $\hat{\Delta}_{1000}(t)$, with a conservative 95\% pointwise confidence intervals for  $\Delta_{1000}(t)$, and an approximate $p$-value of the test of $H_0$ for these comparisons. We see in this example that the plots show clearly that the two models have approximately equal skill at all time point $t$, and further that any discrepancies between the models across $t$ are not significant.  We observed in this case that the empirical rejection rates of $H_0$ for this example tended to be close to, though typically slightly below, the nominal levels of 10\%, 5\%, and 1\%, respectively, for each sample size $N$ used to define the size of the training and testing sets considered, which shows that both the asymptotic result that motivates the test of $H_0$ seems to be reasonably accurate for the sample sizes and examples that we considered, and that the conservative approximations made do not produce a test that is too conservative in practice. We believed that this might be due to the fact that the way that we generate the synthetic game data leads to probabilities $p_i(t)$ that are close to 1/2, in which case the conservative approximation is ineffectual. We also tried adjusting the parameters of the $RS_i$ variable in the simulation so that the home team win rate would be  closer to 90\%, and found that this made the empirical rejection rates somewhat lower, as expected.  We also compared the oracle probability forecasts against the models with noise added, and found that generally the test rejected $H_0$ at a high rate over all significance levels in this case, as expected.

	\par

	\par
\begin{figure} [ht!]
		\centering
		\begin{subfigure}[b]{0.455\textwidth}
			\centering
			\caption[]%
			{{\small}}
			\includegraphics[width=\textwidth]{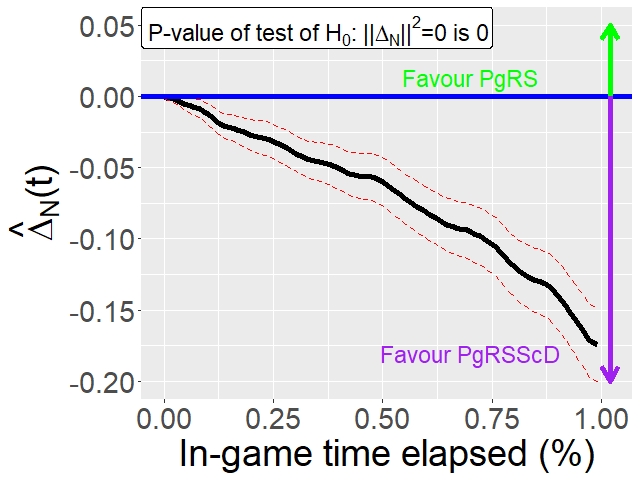}
			\label{fig-sim-RS}
		\end{subfigure}
		\hfill
		\begin{subfigure}[b]{0.455\textwidth}
			\centering
			\caption[]%
			{{\small}}
			\includegraphics[width=\textwidth]{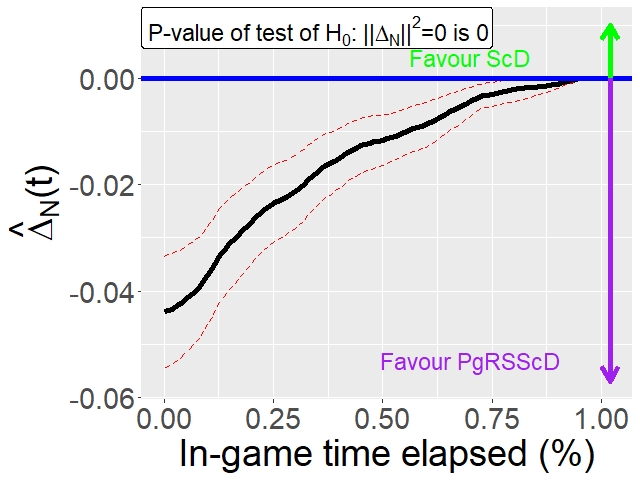}
			\label{fig-sim-SD}
		\end{subfigure}
		\hfill
		\begin{subfigure}[b]{0.455\textwidth}
			\centering
			\caption[]%
			{{\small}}
			\includegraphics[width=\textwidth]{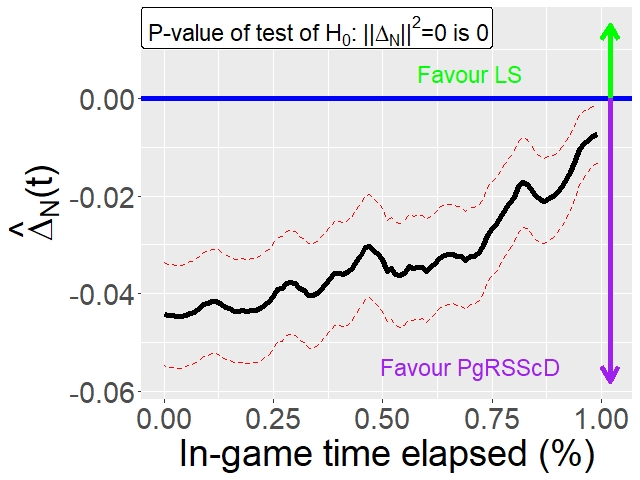}
			\label{fig-sim-LS}
		\end{subfigure}
		\hfill
		\begin{subfigure}[b]{0.455\textwidth}
			\centering
			\caption[]%
			{{\small}}
			\includegraphics[width=\textwidth]{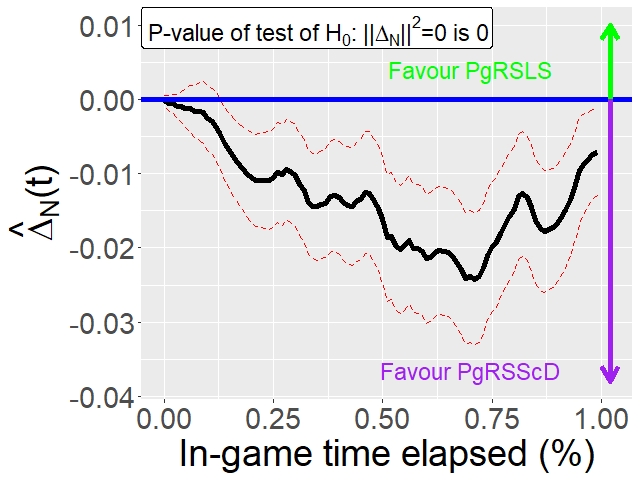}
			\label{fig-sim-RSLS}
		\end{subfigure}
		\caption{ Representative plots of $\hat{\Delta}_{1000}(t)$ derived from simulated data  with 95\% confidence intervals and approximate $p$-values for tests of $H_0$ for (a) {\bf PgRSScD} versus {\bf PgRS} (b) {\bf PgRSScD} versus {\bf ScD} (c) {\bf PgRSScD} versus {\bf LS} (d) {\bf PgRSScD} versus {\bf PgRSLS}.}\label{fig-sim-models}
	\end{figure}

	In terms of comparing the proposed GLM models, similar representative plots comparing the correctly specified model {\bf PgRSScD} to several na\"ive competitors are displayed in Figure \ref{fig-sim-models}. When comparing {\bf PgRSScD} to models that do not adjust their forecasts as the game progresses, our test was always able to distinguish {\bf PgRSScD} as having higher skill.  From Figure \ref{fig-sim-models}, we can see that the expected advantages of {\bf PgRSScD} over the competitor models are transparent in the plot. The relative skill of {\bf PgRSScD} improves over the course of the game compared to models that do not incorporate score information, and for models that do incorporate score information the relative improvement of  {\bf PgRSScD} decays as the game progresses. We do see in Table \ref{tbl-sim-rej} that once the sample size reaches 500 the proposed test of $H_0$ is generally able to distinguish between the correctly specified and na\"ive models with empirical power approaching one.

	\par
	Overall, we found that the proposed test seems to perform well and as expected in many controlled examples, and tends to be conservative. While the test is certainly powerful to differentiate poorly performing models as having low skill compared to competitors, it might struggle to differentiate competitive models without a large sample size ($N\ge 500$ in the examples considered). This along with the graphical tools proposed allow one to easily identify the relative skill of two competing continuously updated probabilistic forecasting models.
	
	\begin{table}
		\centering
		\begin{adjustbox}{max width=\textwidth}
			\begin{tabular}{llllllllll}
				\hline
				\multicolumn{1}{c}{Competing Models}      & \multicolumn{3}{c}{N=100}                                                    & \multicolumn{3}{c}{N=250}                                                    & \multicolumn{3}{c}{N=500}                                                     \\
				\cline{1-1}  \cline{3-3}  \cline{6-6} \cline{9-9}
				\multicolumn{1}{c}{}      & \multicolumn{1}{c}{10\%} & \multicolumn{1}{c}{5\%} & \multicolumn{1}{c}{1\%} & \multicolumn{1}{c}{10\%} & \multicolumn{1}{c}{5\%} & \multicolumn{1}{c}{1\%} & \multicolumn{1}{c}{10\%} & \multicolumn{1}{c}{5\%} & \multicolumn{1}{c}{1\%}  \\
				\hline
				Ora v.s. OraOU        & 0.997                    & 0.993                   & 0.960                   & 1.000                    & 1.000                   & 1.000                   & 1.000                    & 1.000                   & 1.000\\
				Ora v.s. OraBM        & 1.000                    & 0.996                   & 0.963                   & 1.000                    & 1.000                   & 1.000                   & 1.000                    & 1.000                   & 1.000    \\
				OraOU1 v.s OraOU2         & 0.096                    & 0.043                   & 0.007                   & 0.085                    & 0.046                   & 0.005                   & 0.083                    & 0.037                   & 0.002                    \\
				OraBM1 v.s OraBM2         & 0.072                    & 0.028                   & 0.007~                  & 0.081                    & 0.034                   & 0.006                   & 0.089                    & 0.030                   & 0.003                    \\
				PgRSScD v.s. PgRS         & 1.000                    & 0.998                   & 0.972                   & 1.000                    & 1.000                   & 1.000                   & 1.000                    & 1.000                   & 1.000                    \\
				PgRSScD v.s. ScD           & 0.510                    & 0.377                   & 0.176                   & 0.831                    & 0.745                   & 0.509                   & 0.995                    & 0.982                   & 0.907                    \\
				PgRSScD v.s. LS           & 0.795                    & 0.704                   & 0.415                   & 0.990                    & 0.967                   & 0.898                   & 1.000                    & 1.000                   & 1.000                    \\
				PgRSScD v.s. PgRSLS       & 0.820                    & 0.648                   & 0.253                   & 1.000                    & 0.999                   & 0.958                   & 1.000                    & 1.000                   & 1.000                    \\
				\hline
			\end{tabular}
		\end{adjustbox}
		\caption{Empirical rejection rates with nominal levels of 10\%, 5\% and 1\% for the test  $H_0:\|\Delta_N\|^2=0$ in $1000$ independent simulations.}
		\label{tbl-sim-rej}
	\end{table}
	
	\section{Evaluating the skill of ESPN forecasts }\label{sec-app}
	
	\par
	In this section, we apply the methods described in Section \ref{sec3-1} to evaluate the skill of ESPN's continuously updated probabilistic forecasts. In this case all GLM benchmark models are fit using the logit link function.
	
\begin{figure} [ht!]
		\begin{subfigure}[b]{0.455\textwidth}
			\centering
			\caption[]%
			{{\small}}
			\includegraphics[width=\textwidth]{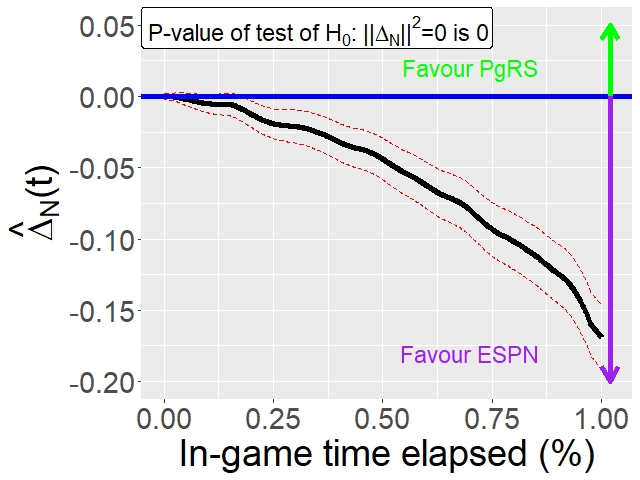}
			\label{fig-loss-PgRS}
		\end{subfigure}
		\hfill
		\begin{subfigure}[b]{0.455\textwidth}
			\centering
			\caption[]%
			{{\small}}
			\includegraphics[width=\textwidth]{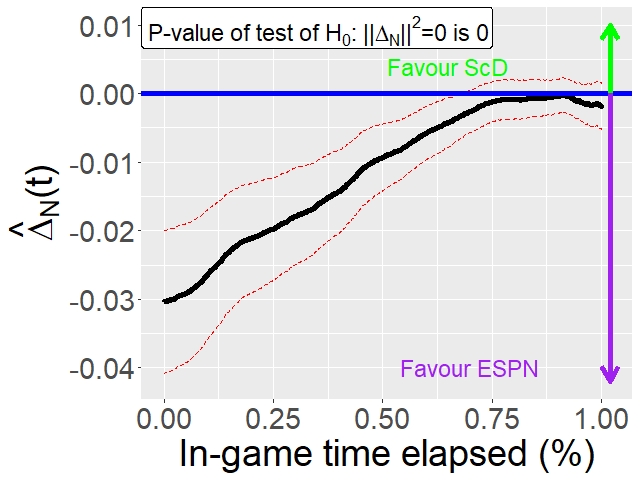}
			\label{fig-loss-SD}
		\end{subfigure}
		\hfill
		\begin{subfigure}[b]{0.455\textwidth}
			\centering
			\caption[]%
			{{\small}}
			\includegraphics[width=\textwidth]{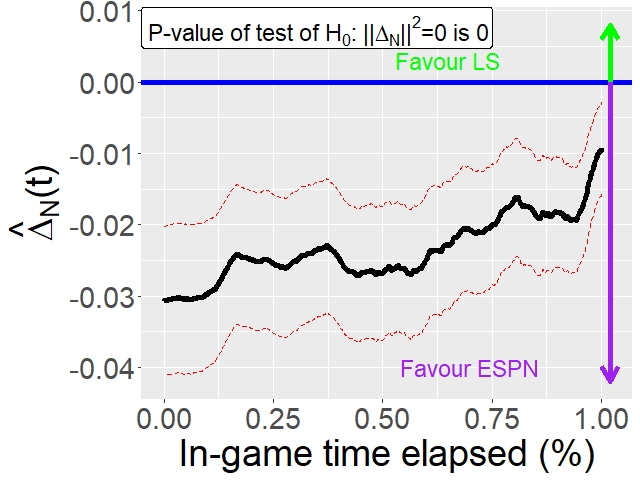}
			\label{fig-loss-LS}
		\end{subfigure}
		\hfill
		\begin{subfigure}[b]{0.455\textwidth}
			\centering
			\caption[]%
			{{\small}}
			\includegraphics[width=\textwidth]{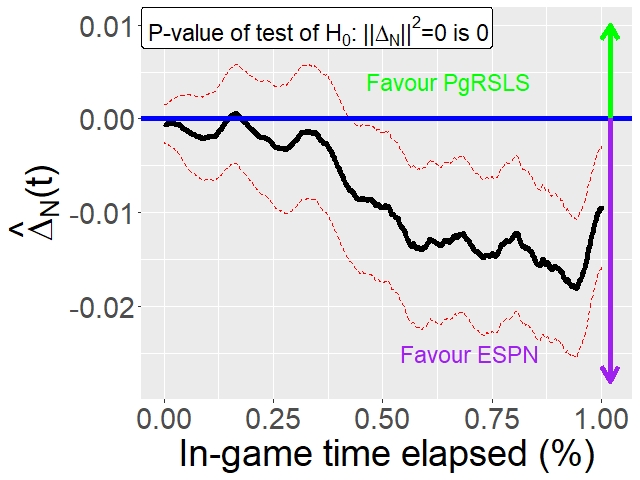}
			\label{fig-loss-PgRSLS}
		\end{subfigure}
		\caption{ Plots of $\hat{\Delta}_{1213}(t)$ based on the 2018-2019 season forecasts with 95\% confidence intervals and approximate $p$-values for tests of $H_0$ for (a) {\bf ESPN} versus {\bf PgRS}  (b) {\bf ESPN} versus  {\bf ScD}  (c) {\bf ESPN} versus {\bf LS}  (d) {\bf ESPN} versus {\bf PgRSLS}.}\label{fig-loss-models}
	\end{figure}

	Figure \ref{fig-loss-models} shows plots $\hat{\Delta}_{1213}(t)$ with conservative 95\% confidence intervals, as well as approximate $p$-values of the test of $H_0$ for comparisons of the ESPN forecasts with the na\"ive models {\bf PgRS},  {\bf ScD}, {\bf LS}, and {\bf PgRSLS}. These plots suggest that, in aggregate, the ESPN forecasts significantly out perform these models. The specific points in the game at which the ESPN forecasts exhibit higher skill compared to these benchmarks is also clear in the plots. For the models that use relative team strength as encoded by ESPN's initial home win probability as a covariate, the relative skill is similar to ESPN's forecasts early in the game, and similarly those that make use of the score difference improve relative to the ESPN forecasts towards the end of the game. For example, the model based on the score difference alone is strongly outperformed by the ESPN forecasts early in the game, but their forecasts have indistinguishable skill towards the end of the game.

	\par
\begin{figure} [ht]
		\centering
		{{\small}} \label{fig-logit-espn}
		\includegraphics[width=0.72\textwidth]{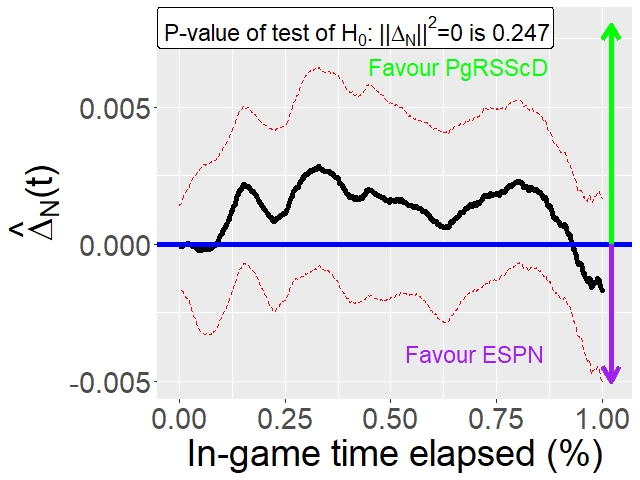}
		\caption{ Plots of $\hat{\Delta}_{1213}(t)$ based on the 2018-2019 season forecasts with 95\% confidence interval and approximate $p$-value for tests of $H_0$ for {\bf ESPN} versus {\bf PgRSScD}} \label{fig-quality}.
	\end{figure}

	We also compared the ESPN forecasts to those of the somewhat less na\"ive model {\bf PgRSScD} using a logit link.  Figure \ref{fig-quality} shows a plot $\hat{\Delta}_{1213}(t)$ based on the 2018-2019 season with conservative 95\% confidence intervals, as well as approximate $p$-values of the test of $H_0$ for this comparison. In absolute terms, the estimated skill as measured by the Brier score generally favoured the simple logit model {\bf PgRSScD}, with the exception of the last moments in the game. However, we do see from this analysis that the difference is apparently not statistically significant at the 5\% level at any game time point based on the conservative confidence interval estimates, nor is it significant when the difference is aggregated across time points. We found it interesting that the ESPN's sophisticated proprietary model, which ostensibly makes use of more nuanced information about the game status and more sophisticated models, did not significantly outperform a simple Logistic regression model. One might draw the conclusion based on this that whatever additional information used by ESPN's model in producing these forecasts is not clearly beneficial for the purpose of forecasting, except for in the final moments of the game.

	\section{Discussion}
	\par
	Motivated by evaluating forecasts of NBA basketball games, we have developed graphical tools and statistical tests for assessing the calibration and relative skill of continuously updated probabilistic forecasts. These were studied via a simulation study of synthetic ``basketball games", and applied to evaluating and comparing the forecasts published on ESPN and a number of competing models. In terms of calibration, the ESPN forecasts, as well as forecasts produced from simple logistic regression models using the in-game score difference and/or pre-game relative strength of teams as covariates, appear reasonably well-calibrated. In terms of skill, the ESPN forecasts exhibited significantly higher skill over na\"ive models, but did not demonstrate superiority over simple logistic regression models based on the score difference and relative team strength.

	We conclude with a few remarks about some ideas that we considered but chose not to include in the paper, and avenues for future work. It is noteworthy that the confidence intervals defined in \eqref{eq-conf-int-diff} may be made narrower and less conservative by using auxiliary information to improve the approximation of replacing $p_i(t)(1-p_i(t))$ with the upper bound $1/4$. We considered a number of methods to achieve this, including employing auxiliary models and covariates to estimate $p_i(t)(1-p_i(t))$ solely in the variance estimation step, but found both that this changed the intervals little, and lead to poor performance near the end of the game. For a discussion of similar methods, see Sections 3.4 and 3.5 in \citep{lai-2011}.
	
	\par
	The basic reason why nested models often cannot be compared by means of the confidence intervals introduced and the test of $H_0$ derived using Theorem \ref{thm-test} is that nested models may lead to a covariance kernel $\hat{C}_{cons}$ that is approximately (and asymptotically) degenerate. It might be interesting to adapt this result, and the corresponding intervals and test, to this case, for instance using methods similar to those described in \cite{clark-2015}.

	\par
	
	Lastly, we arrived after experimenting with many ways of constructing Figure \ref{fig-veri-plot} in Section \ref{sec-calib} and the calibration plots in Figure \ref{fig-mmplots} in Section \ref{sec3-1} on using the Wilson interval with a bin-size of $10\%$. There are many other reasonable candidates to construct such intervals, including the typical normal approximation interval, and the Agresti and Coull interval \citep{agresti-1998}, among several others, and we found that these performed similarly well with differences only appearing in the more extremal (with centers closer to 0 or 1) bins. Data driven approaches for deciding on an interval type as well as bin-size for these plots is worthy of further study.

	\spacingset{1.00} 
	
	\bibliographystyle{apalike}
	\bibliography{NBAFinalSUBMIT}
	
	\section*{Appendix}
	\appendix
	\section{Definition of $p_i(t)$ and proof of Theorem \ref{thm-test}} \label{sec-appendix}
	{\small
		
		In order to formally define $p_i(t)$, we first define two collections of sigma-algebras: $\mathcal{F}_i$, which is the information available up to and including the $i'th$ game, and $\mathcal{F}_{i-1,t}$ which is the information in all games up to and including the $i-1$'st game, along with the information up to time $t$ in the $i'th$ game. Clearly then for all $t\in [0,1]$, $\mathcal{F}_{i-1,t} \subset \F_i \subset \F_{i,t}$. Let $Y_i$ denote the zero-one variables encoding wins for the home team. We define $p_i(t) = E[Y_i | \F_{i-1,t}]$ to be the probability that the home team wins in the $i$'th game given the information up to time $t$ in that game (and previous games), which we are aiming to forecast.
		
		Suppose that we have two methods $\hat{p}^A_i(t)$, $\hat{p}^B_i(t)$ for forecasting the in-game probabilities of the home team winning in game $i$ at time $t$.
		
		\begin{assumption}\label{ass-meas}
			$\hat{p}^A_i(t)$, $\hat{p}^B_i(t)$ are both measurable with respect to $\mathcal{F}_{i-1,t}$. In other words, only the information from all previous games as well as the current game up to time $t$ are used to produce the forecasts $\hat{p}^A_i(t)$, $\hat{p}^B_i(t)$.
		\end{assumption}
		
		Let $L^2[0,1]$ denote the space of square integrable functions defined on $[0,1]$, which has canonical inner-product defined for $f,g\in L^2[0,1]$ as $\langle f,g \rangle = \int_{0}^{1}f(t)g(t)dt$. We assume here that $L(a,b)=(a-b)^2$ is the Brier loss, but the following results generalize to any loss functions with a linear equivalent \citep{lai-2011}. Define
		
		\begin{align*}
			Z_N(t) &= \sqrt{N}[\hat{\Delta}_N(t)-\Delta_N(t)] \\&= \frac{1}{\sqrt{N}} \sum_{i=1}^{N} \left[L(Y_i,\hat{p}^A_i(t)) -L(Y_i,\hat{p}^B_i(t))-[L(p_i(t),\hat{p}^A_i(t)) -L(p_i(t),\hat{p}^B_i(t))]\right].
		\end{align*}
		Note that under the null hypothesis $H_0: \; \|\Delta_N\|^2=0$ of equivalent forecasting skill between the models $A$ and $B$, $Z_N(t) = \sqrt{N}\hat{\Delta}_N(t)$ in $L^2$ sense. In order to formally state and prove Theorem \ref{thm-test}, we also make the following assumptions. A sequence of covariance kernels that arises in calculating the distribution of $Z_N(t)$ is
		
		\begin{align}\label{cndef}
			C_N(t,s) &= \frac{4}{N} \sum_{i=1}^{N} \left( E[(\hat{p}^A_i(s)-\hat{p}^B_i(s))(\hat{p}^A_i(t)-\hat{p}^B_i(t)) p_i(s)(1-p_i(s))  | \F_{i-1}]  \mathds{1}_{s>t}  \right. \notag \\
			&+ \left. E[(\hat{p}^A_i(s)-\hat{p}^B_i(s))(\hat{p}^A_i(t)-\hat{p}^B_i(t)) p_i(t)(1-p_i(t))  | \F_{i-1}]  \mathds{1}_{t>s} \right),
		\end{align}
		
		where $\mathds{1}$ is the indicator function. As we will show, $C_N$ is essentially the average covariance function of the (centered) difference between the forecasts $\hat{p}_i^A(t)-\hat{p}_i^B(t)$. This kernel is evidently symmetric for each $N$, and we show below that it is also positive (semi-)definite, which is to say that for all functions $f\in L^2[0,1]$,
		
		$$
		\int_{0}^{1} \hspace{-.1cm} \int_{0}^{1} C_N(t,s)f(t)f(s)dtds \ge 0.
		$$
		
		\begin{assumption}\label{ass-degen} There exists a symmetric, positive definite kernel $C$ satisfying $\int_{0}^{1}C(t,t)dt < \infty$ such that
			\begin{align}\label{cnconv}
				\int_{0}^{1} \hspace{-.1cm} \int_{0}^{1} [C_N(t,s)-C(t,s)]^2 dtds =o_P(1).
			\end{align}
		\end{assumption}
		
		Assumption \ref{ass-degen} basically entails that there is a degree of ``stationarity" or ``ergodicity" between subsequent forecasts. This would be implied if, for example, subsequent forecasts in each game were independent of each other and exhibit an approximately stable distribution, although independence could be relaxed a great deal here with \eqref{cnconv} still holding. This condition is analogous to condition (a) in Theorem 1 of \cite{dawid-1993}, the validity of which in probabilistic forecast evaluation is discussed there, and this condition is also central to the results of \cite{lai-2011}.
		
		The kernel $C$ defined in Assumption \ref{ass-degen} may be used to define a symmetric, positive definite linear operator $\mathcal{C}: L^2[0,1] \mapsto L^2[0,1]$ via
		
		\begin{align}\label{copdef}
			\mathcal{C}(f)(t) = \int_{0}^{1} C(t,s)f(s)ds,
		\end{align}
		which by Mercer's Theorem \citep{bosq:2000} must have associated to it a decreasing sequence of eigenvalues $\lambda_1 \ge \lambda_2 \ge \cdots$ and corresponding orthonormal eigenfunctions $\phi_1,\phi_2,\ldots \in L^2[0,1]$ satisfying $\mathcal{C}(\phi_i) = \lambda_i \phi_i$. A similar collection of eigenvalues and $\lambda_{1,N} \ge \lambda_{2,N} \ge \cdots$ may be defined the operator $\mathcal{C}_N$ obtained by replacing $C$ by $C_N$ in the definition \eqref{copdef}.
		
		\begin{assumption}\label{ass-tight}
			The sequence of variables $Z_N$ are uniformly tight in $L^2[0,1]$ (see pg. 46 of \cite{bosq:2000}).
		\end{assumption}
		Uniform tightness is used in order to extend asymptotic Gaussianity of the finite dimensional distributions of the process $Z_N$ to the entire process. Intuitively assuming tightness in this context assumes a degree of continuity or ``smoothness" to the forecasts $\hat{p}_i^A(t)$ and $\hat{p}_i^B(t)$ as a function of $t$. A sufficient condition to imply Assumption \ref{ass-tight} is
		
		$$
		\sup_{N\ge 1} \sum_{i=1}^{\infty} \lambda_{i,N} < \infty,
		$$
		see the proof on page 52 of \cite{bosq:2000}.
		
		{\it Restatement of Theorem \ref{thm-test}} Under $H_0$ and Assumptions \ref{ass-meas}-\ref{ass-tight},
		
		$$
		\|Z_N\|^2 \stackrel{D}{\to} \sum_{i=1}^{\infty} \lambda_i \chi_i^2(1),
		$$
		where the $\chi_i^2(1)$, $i=1,2,\ldots$ are independent and identically distributed $\chi^2$ random variables with one degree of freedom and the constants $\{\lambda_i, \; i\ge 1\}$ are defined in \eqref{copdef}. These constants may be asymptotically conservatively estimated by the eigenvalues of
		$$
		\hat{C}_{cons}(t,s) = \frac{1}{N} \sum_{i=1}^{N} [\hat{p}^A_i(t)-\hat{p}^B_i(t)][\hat{p}^A_i(s)-\hat{p}^B_i(s)].
		$$

		\begin{proof}
			
			For all test functions $v\in L^2[0,1]$, and using the definition of the Brier score, we have that
			
			\begin{align*}
				\langle Z_N, v \rangle/2 &= \sqrt{N}[\hat{\Delta}_N(t)-\Delta_N(t)]/2 = \frac{1}{2\sqrt{n}} \sum_{i=1}^{N} \langle L(Y_i,\hat{p}^A_i) -L(Y_i,\hat{p}^B_i)-[L(p_i,\hat{p}^A_i) -L(p_i,\hat{p}^B_i)],v\rangle \\
				&=  \frac{1}{\sqrt{N}} \sum_{i=1}^{N} \langle (\hat{p}^A_i-\hat{p}^B_i)(p_i-Y_i) ,v\rangle =: \frac{1}{\sqrt{N}} \sum_{i=1}^{N} X_{i,v}.
			\end{align*}
			$X_{i,v}$ is a martingale difference sequence with respect to the filtration $\F_i$. To see this, notice by the tower property of conditional expectation and Fubini's theorem that
			
			$$
			E[X_{i,v} | \F_{i-1} ] = E[ E[ X_{i,v} |\F_{i-1,t}]\F_{i-1}]= E[  \langle E [(\hat{p}^A_i-\hat{p}^B_i)(p_i-Y_i)|\F_{i-1,t}] ,v\rangle |\F_{i-1}]=0,
			$$
			since $E [(\hat{p}^A_i(t)-\hat{p}^B_i(t))(p_i(t)-Y_i)|\F_{i-1,t}]=(\hat{p}^A_i(t)-\hat{p}^B_i(t))E[(p_i(t)-Y_i)|\F_{i-1,t}]=0$, a.s. for all $t$, using Assumption \ref{ass-meas}. Further,
			\begin{align}\label{cond-eq}
				E[X_{i,v}^2 | \F_{i-1} ] = \int_{0}^{1} \hspace{-.1cm} \int_{0}^{1}  E[(\hat{p}^A_i(t)-\hat{p}^B_i(t))(p_i(t)-Y_i)(\hat{p}^A_i(s)-\hat{p}^B_i(s))(p_i(s)-Y_i)v(t)v(s) | \F_{i-1}] dtds
			\end{align}
			Suppose $s>t$, then using the tower property
			\begin{align*}
				E[(\hat{p}^A_i(s)-&\hat{p}^B_i(s))(p_i(s)- Y_i)(\hat{p}^A_i(t)-\hat{p}^B_i(t))(p_i(t)-Y_i)v(t)v(s) | \F_{i-1}]  \\
				&=  E[E[(\hat{p}^A_i(s)-\hat{p}^B_i(s))(p_i(s)-Y_i)(\hat{p}^A_i(t)-\hat{p}^B_i(t))(p_i(t)-Y_i)v(t)v(s)| \F_{i-1,s}] | \F_{i-1}] \\
				&= E[(\hat{p}^A_i(s)-\hat{p}^B_i(s))(\hat{p}^A_i(t)-\hat{p}^B_i(t)) v(t)v(s) E[(p_i(s)-Y_i)(p_i(t)-Y_i)| \F_{i-1,s}] | \F_{i-1}].
			\end{align*}
			Note that $Y_i| \F_{i-1,s}$ is a Bernoulli($p_i(s)$) random variable, it follows that $ E[(p_i(s)-Y_i)(p_i(t)-Y_i)| \F_{i-1,s}] = p_i(s)(1-p_i(s))$. From this we obtain that for $s>t$,
			
			\begin{align*}
				E[(\hat{p}^A_i(s)-\hat{p}^B_i(s))(p_i(s)-& Y_i)(\hat{p}^A_i(t)-\hat{p}^B_i(t))(p_i(t)-Y_i)v(t)v(s) | \F_{i-1}]  \\
				&= E[(\hat{p}^A_i(s)-\hat{p}^B_i(s))(\hat{p}^A_i(t)-\hat{p}^B_i(t)) p_i(s)(1-p_i(s))  | \F_{i-1}] v(t)v(s).
			\end{align*}
			Plugging into \eqref{cond-eq} and applying the same calculation when $t>s$ gives that
			\begin{align*}
				E[X_{i,v}^2 | \F_{i-1} ]
				&= \int_{0}^{1} \hspace{-.1cm} \int_{0}^{1} \{ E[(\hat{p}^A_i(s)-\hat{p}^B_i(s))(\hat{p}^A_i(t)-\hat{p}^B_i(t)) p_i(s)(1-p_i(s))  | \F_{i-1}]  \mathds{1}_{s>t}  \\
				&+ E[(\hat{p}^A_i(s)-\hat{p}^B_i(s))(\hat{p}^A_i(t)-\hat{p}^B_i(t)) p_i(t)(1-p_i(t))  | \F_{i-1}]  \mathds{1}_{t>s} \} v(t)v(s) dtds. \\
			\end{align*}
			It follows that
			
			$$
			\frac{1}{N} \sum_{i=1}^{N} E[X_{i,v}^2 | \F_{i-1} ] =  \int_{0}^{1} \hspace{-.1cm} \int_{0}^{1} \frac{1}{4}C_N(t,s)v(t)v(s)dtds. 
			$$
			Hence using Assumption \ref{ass-degen},
			\begin{align}\label{condconv1}
				\frac{1}{N} \sum_{i=1}^{N} E[X_{i,v}^2 | \F_{i-1} ] \stackrel{P}{\to} \int_{0}^{1} \hspace{-.1cm} \int_{0}^{1} \frac{1}{4}C(t,s)v(t)v(s)dtds = \langle \mathcal{C}(v),v \rangle =: \sigma_v^2.
			\end{align}
			
			We consider the cases of $\sigma_v^2=0$ and $\sigma_v^2 > 0$ separately. In the case where $\sigma_v^2 = 0$, we first note that, using the Cauchy-Schwarz inequality, $E[X_{i,v}^2 | \F_{i-1} ] \le \|v\|^2  E[ \|(\hat{p}^A_i(\cdot)-\hat{p}^B_i(\cdot))(p_i(\cdot)-Y_i)\|^2 |\F_{i-1} ]$, which is uniformly bounded, and hence $1/N \sum_{i=1}^{N} E[X_{i,v}^2 | \F_{i-1} ]$  is uniformly integrable. It follows then from \eqref{condconv1} that $\sigma_v^2 = 0$ implies
			
			$$
			\mbox{Var}\left( \frac{1}{\sqrt{N}} \sum_{i=1}^{N} X_{i,v} \right) = E\left\{ \frac{1}{N} \sum_{i=1}^{N} E[X_{i,v}^2 | \F_{i-1} ] \right\} \to 0.
			$$
			
			When $\sigma_v^2 > 0$, let $S_{N,v}^2 = \mbox{Var}(\sum_{i=1}^{N} X_{i,v}) =  \sum_{i=1}^{N} E[X_{i,v}^2 ]$. Using \eqref{condconv1}, it follows that there exists a constant $D>0$ so that for all $N$ sufficiently large, $S_{N,v}^2 \ge DN$. Thus, for all $\epsilon>0$ and $N$ sufficiently large, we have due to the uniform boundedness of $X_{i,v}^2$ that
			
			$$
			\frac{1}{S_{N,v}^2}  \sum_{i=1}^{N} E X_{i,v}^2 \mathds{1}_{\{ |X_{i,v}| \ge \epsilon S_{N,v} \}} = 0.
			$$
			Hence both the Lindeberg condition and the asymptotic constancy of the conditional variance condition of the Martingale central limit theorem, c.f. Theorem 1 of \cite{brown1971}, hold, giving that $\langle Z_N, v \rangle/2 \stackrel{D}{\to} \mathcal{N}(0,\sigma_v^2)$. This shows that the finite dimensional distributions of $Z_N$ asymptotically coincide with those of a Gaussian process with covariance kernel $C$. This along with Assumption \ref{ass-tight} imply that $Z_N$ converges weakly in $L^2[0,1]$ to a Gaussian process with covariance kernel $C$ (see Prohorov's theorem, Theorem 2.6 in \cite{bosq:2000}). The form of the limiting distribution of $\|Z_N\|^2$ follows from the continuous mapping theorem and the Karhunen-Lo\'eve representation theorem.
			
			The fact that the eigenvalues of the kernel $\hat{C}_{cons}$ are uniformly larger than those of $C_N$ follows since $\hat{C}_{cons}$ is obtained by replacing  $p_i(t)(1-p_i(t))$ and $p_i(s)(1-p_i(s))$ with the upper bound $1/4$.
		\end{proof}
	}
	
\end{document}